\documentclass[preprint,aps,superscriptaddress]{revtex4}
\usepackage{amssymb}
\usepackage{amsmath}
\usepackage{graphicx}
\usepackage{color}

\newcommand\beq{ \begin{eqnarray} }
\newcommand\eeq{ \end{eqnarray} }

\begin{document}

\title{Shear Viscosity of a Gluon Plasma in Perturbative QCD}
\author{Jiunn-Wei Chen}
\affiliation{Department of Physics and Center for Theoretical Sciences, National Taiwan
University, Taipei 10617, Taiwan}
\author{Hui Dong}
\affiliation{School of Physics, Shandong University, Shandong 250100, People's Republic
of China}
\author{Kazuaki Ohnishi}
\affiliation{Department of Physics and Center for Theoretical Sciences, National Taiwan
University, Taipei 10617, Taiwan}
\author{Qun Wang}
\affiliation{Interdisciplinary Center for Theoretical Study and Department of Modern
Physics, University of Science and Technology of China, Anhui 230026,
People's Republic of China}

\begin{abstract}
We calculate the shear viscosity $\left( \eta \right) $ to entropy density $%
(s)$ ratio $\eta /s$ of a gluon plasma in kinetic theory including the $%
gg\rightarrow gg$ and $gg\rightarrow ggg$ processes. Due to the suppressed
contribution to $\eta $\ in the $gg\rightarrow gg$ forward scattering, it is
known that the gluon bremsstrahlung $gg\leftrightarrow ggg$ process also
contributes at the same order ($O(\alpha _{s}^{-2})$) in perturbative QCD.
Using the Gunion-Bertsch formula for the $gg\rightarrow ggg$\ matrix element
which is valid for the limit of soft bremsstrahlung, we find that the result
is sensitive to whether the same limit is taken for the phase space. Using
the exact phase space, the $gg\leftrightarrow ggg$\ contribution becomes
more important to $\eta $\ than $gg\rightarrow gg$\ for $\alpha _{s}\gtrsim
2\times 10^{-3}$.\ Therefore, at $\alpha _{s}=0.1$, $\eta /s\simeq 1.0$,
between 2.7 obtained by Arnold, Moore and Yaffe (AMY) and 0.5 obtained by Xu
and Greiner. If the soft bremsstrahlung limit is imposed on the phase space
such that the recoil effect from the bremsstrahlung gluon is neglected, then
the correction from the $gg\leftrightarrow ggg$\ process is about 10-30\% of
the total which is close to AMY's prediction. This shows that the soft
bremsstrahlung approximation is not as good as previously expected.
\end{abstract}

\maketitle

\begin{flushright}
USTC-ICTS-09-10
\end{flushright}

%\date{\today}

\section{Introduction}

One of the most surprising discoveries at the Relativistic Heavy Ion
Collider (RHIC) is that the hot and dense matter (believed to be a quark
gluon plasma (QGP), see \cite%
{Gyulassy:2004zy,Shuryak:2004cy,Stoecker:2004qu,Jacobs:2004qv} for reviews)
formed in collisions appears to be a near-perfect fluid \cite%
{RHIC,Huovinen:2001cy,Molnar:2001ux,Teaney:2000cw,Hirano:2002ds,
Teaney:2003pb,Muronga:2004sf,Heinz:2005bw,Romatschke:2007mq}. The remanent
of the non-central collisions shows collective motion (elliptic flow) with a
shear viscosity ($\eta $) to entropy density ($s$) ratio $\eta /s=0.1\pm 0.1(%
\mathrm{theory})\pm 0.08(\mathrm{experiment})$ \cite{Luzum:2008cw}. This $%
\eta /s$ ratio is close to a conjectured minimum bound $1/4\pi $ \cite%
{Kovtun:2004de}, which is motivated by uncertainty principle \cite%
{Danielewicz:1984ww} and gauge/string duality \cite%
{Policastro:2001yc,Policastro:2002se,Herzog:2002fn,Buchel:2003tz}. Since
smaller $\eta /s$ implies stronger particle interactions, contrary to the
conventional picture, the QGP produced at RHIC tends to be a strongly
interacting fluid instead of a weakly interacting gas.

However, a recent perturbative QCD calculation\ of $\eta /s$ of a gluon
plasma by Xu and Greiner (XG) \cite{Xu:2007ns} indicates that the gluon
elastic scattering $gg\rightarrow gg$ does not give the dominant
contribution. They found that $\eta /s$ for the gluon bremsstrahlung process
$gg\leftrightarrow ggg$ is about 1/7 of that for $gg\rightarrow gg$, which
means the contribution to the shear viscosity from $gg\leftrightarrow ggg$
is 7 times as important as that from $gg\rightarrow gg$. This would bring $%
\eta /s$ down to $1/4\pi $ when strong coupling constant $\alpha
_{s}\simeq 0.6$. This implies that the near-perfect QGP might still
be described by perturbative QCD and that the conventional picture
could still be valid. Their conclusion is quite different from an
earlier study by Arnold, Moore and Yaffe (AMY) \cite{Arnold:2003zc}
(for a recent review, see, e.g., \cite{Arnold:2007pg}). AMY found
that $gg\leftrightarrow ggg$ only contributes at $10\%$ level for
the three flavor quark diffusion constant for $\alpha
_{s}<0.3$. For comparison, XG have $\eta /s\simeq 0.5$\ at $\alpha _{s}=0.1$%
, while AMY have $\eta /s\simeq 2.7$\ (note that only $\eta $\ was computed
in \cite{Arnold:2003zc}, the free gluon $s$\ is inserted by us for
comparison).

Both approaches of XG and AMY are based on kinetic theory. However, the main
points of differences are: 1) A parton cascade model \cite{Xu:2004mz} is
used by XG to solve the Boltzmann equation. Since the bosonic nature of
gluons is hard to implement in real time simulations in this model, gluons
are treated as a Boltzmann gas (i.e. a classical gas). For AMY, the
Boltzmann equation is solved in a variation method without taking the
Boltzmann gas approximation. 2) The $Ng\leftrightarrow (N+1)g$ processes, $%
N=2,3,4\ldots $, are approximated by the effective $g\leftrightarrow gg$
splitting in AMY where the two gluons are nearly collinear with a small
splitting angle, while the $gg\leftrightarrow ggg$\ process is used in XG
where the bremsstrahlung gluon is soft but it can have a large splitting
angle with its mother gluon. More specifically, in XG, the Gunion-Bertsch
formula \cite{Gunion:1981qs} for the $gg\rightarrow ggg$\ matrix element
squared in Eq. (\ref{eq:matrix-e}) is used. This formula is valid for the
limit of soft but not necessarily collinear gluon bremsstrahlung. For the
phase space, XG uses the exact phase space for the three gluon
configurations (called \textquotedblleft three-body-like\textquotedblright\
phase space in this paper). In principle, if the soft bremsstrahlung limit
is a good approximation of the $gg\rightarrow ggg$ process, one should be
able to impose the same limit to the phase space as well and get
approximately the same result. In this limit, the recoil effect from the
bremsstrahlung gluon is neglected, and the phase space (for the two hard
gluons) is called \textquotedblleft two-body-like\textquotedblright\ here.

In this paper, we will perform a third independent calculation for
comparison. We will use the same inputs on the Gunion-Bertsch formula for
the gluon scattering amplitudes (modulo a factor 2 in Eq. (\ref{eq:matrix-e}%
)) with the soft gluon bremsstrahlung approximation as XG but we will solve
the Boltzmann equation variationally as AMY without taking the Boltzmann gas
approximation. We will also test the robustness of the soft gluon
bremsstrahlung approximation by comparing the results with the two- and
three-body-like phase space.

\section{Kinetic Theory}

Using the Kubo formula, $\eta $ can be calculated through the linearized
response function of a thermal equilibrium state
\begin{equation}
\eta =-\frac{1}{5}\int_{-\infty }^{0}\mathrm{d}t^{\prime }\int_{-\infty
}^{t^{\prime }}\mathrm{d}t\int \mathrm{d}x^{3}\langle \left[
T^{ij}(0),T^{ij}(\mathbf{x},t)\right] \rangle ,
\end{equation}%
where $T^{ij}$ is the spatial part of the off-diagonal energy momentum
tensor. In a leading order (LO) expansion of the coupling constant, there
are an infinite number of diagrams \cite{Jeon}. However, it is proven that
the summation of the LO diagrams in a weakly coupled $\phi ^{4}$ theory \cite%
{Jeon} or in hot QED \cite{Gagnon:2007qt} is equivalent to solving the
linearized Boltzmann equation with temperature-dependent particle masses and
scattering amplitudes. The conclusion is expected to hold in weakly coupled
systems and can as well be used to compute the LO transport coefficients in
QCD-like theories \cite{Arnold:2000dr,Arnold:2003zc}, hadronic gases \cite%
{Prakash:1993bt,Dobado:2003wr,Dobado:2001jf,Chen:2006iga,Chen:2007xe,Itakura:2007mx}
and weakly coupled scalar field theories \cite{Jeon,Moore:2007ib,Chen:2007jq}%
.

The Boltzmann equation of a hot gluon plasma describes the evolution of the
color averaged gluon distribution function $f=f(x,p,t)\equiv f_{p}(x)$ (a
function of space, time and momentum) as \cite%
{Heinz:1984yq,Elze:1986qd,Blaizot:1999xk,Baier:2000sb,Wang:2001dm}
\begin{equation}
\frac{p^{\mu }}{E_{p}}\partial _{\mu }f_{p}=C[f],  \label{Boltzmann}
\end{equation}%
where $E_{p}=p$ for massless gluons. The driving force for the evolution is
the particle scattering in the microscopic theory described by the collision
term $C$ which is a functional of $f$. It is known that, to compute $\eta $
to LO in the coupling constant $\alpha _{s}$, we need to include $%
gg\rightarrow gg$ and $gg\leftrightarrow ggg$ scattering in $C$ \cite%
{Baier:2000sb,Arnold:2002zm}. We will show this more explicitly later. In
thermal equilibrium, the gluon distribution $f_{p}^{0}$ is static,
homogeneous and isotropic and hence $\partial _{\mu }f_{p}^{0}=0$ at every $%
x $. This implies $C[f^{0}]=0$ or detailed balance whose solution is just
the Bose-Einstein distribution function $f_{p}^{0}=1/(e^{E_{p}/T}-1)$. When
the system is not in thermal equilibrium, there will be momentum flow due to
the breakdown of detailed balance. The momentum flow can be characterized by
a velocity field $V(x)$. The deviation from thermal equilibrium can be
characterized by the inhomogeneity of $V(x)$ or the derivative expansions of
$V(x)$. For simplicity, we work in the comoving frame of the fluid element
at point $x$ with $V=0$ and to the order of first derivatives of $V$. Thus
the distribution function can be parametrized as%
\begin{equation}
f_{p}=f_{p}^{0}[1-\chi _{p}(1+f_{p}^{0})],  \label{fp}
\end{equation}%
where
\begin{equation}
\chi _{p}=\left[ A(p)\nabla \cdot \mathbf{V}+B(p)\hat{p}_{[i}\hat{p}%
_{j]}\nabla _{\lbrack i}V_{j]}\right] /T\ ,  \label{eq:chi}
\end{equation}%
and where the symmetric traceless combinations $\hat{p}_{[i}\hat{p}_{j]}=%
\hat{p}^{i}\hat{p}^{j}-\delta _{ij}/3$ and $\nabla _{\lbrack i}V_{j]}=\left(
\nabla _{i}V_{j}+\nabla _{j}V_{i}\right) /2-\nabla \cdot V\delta _{ij}/3$.
Note that the time derivatives do not appear because they can be related to
the spatial derivatives by virtue of the conservation of energy momentum
tensor. Analogously the deviation of the energy momentum tensor away from
its equilibrium value can be parametrized by the bulk ($\zeta $) and shear ($%
\eta $) viscosities
\begin{equation}
\delta T_{ij}=\zeta \delta _{ij}\nabla \cdot \mathbf{V}-2\eta \nabla
_{\lbrack i}V_{j]}\ .
\end{equation}%
Using the definition in kinetic theory $T_{\mu \nu }=N_{g}\int \frac{d^{3}%
\mathbf{p}}{(2\pi )^{3}}\frac{p_{\mu }p_{\nu }}{E_{p}}f_{p}(x)$, one obtains
\begin{equation}
\eta =\frac{N_{g}\beta }{15}\int \frac{\mathrm{d}^{3}\mathbf{p}}{(2\pi )^{3}}%
\frac{p^{2}}{E_{p}}f_{p}^{0}\left( 1+f_{p}^{0}\right) B(p)\ ,  \label{eta1}
\end{equation}%
where $N_{g}=16$ is the gluon polarization and color degeneracy.

Following the standard procedure (see e.g. \cite{Jeon}) and making use of
the Boltzmann equation satisfied by $B(p)$, Eq.(\ref{eta1}) can be recast
into
\begin{eqnarray}
\eta &=&\frac{N_{g}^{2}\beta }{80}\int \prod_{i=1}^{4}\frac{d^{3}\mathbf{p}%
_{i}}{(2\pi )^{3}2E_{i}}|M_{12\rightarrow 34}|^{2}(2\pi )^{4}\delta
^{4}(p_{1}+p_{2}-p_{3}-p_{4})(1+f_{1}^{0})(1+f_{2}^{0})f_{3}^{0}f_{4}^{0}
\notag \\
&&\times \lbrack B_{ij}(p_{4})+B_{ij}(p_{3})-B_{ij}(p_{2})-B_{ij}(p_{1})]^{2}
\notag \\
&&+\frac{N_{g}^{2}\beta }{120}\int \prod_{i=1}^{5}\frac{d^{3}\mathbf{p}_{i}}{%
(2\pi )^{3}2E_{i}}|M_{12\rightarrow 345}|^{2}(2\pi )^{4}\delta
^{4}(p_{1}+p_{2}-p_{3}-p_{4}-p_{5})(1+f_{1}^{0})(1+f_{2}^{0})f_{3}^{0}f_{4}^{0}f_{5}^{0}
\notag \\
&&\times \lbrack
B_{ij}(p_{5})+B_{ij}(p_{4})+B_{ij}(p_{3})-B_{ij}(p_{2})-B_{ij}(p_{1})]^{2},
\label{eta2}
\end{eqnarray}
where $B_{ij}(p)\equiv B(p)(\hat{p}^{i}\hat{p}^{j}-\frac{1}{3}\delta _{ij})$
and $M_{12\rightarrow 34}$ and $M_{12\rightarrow 345}$ are amplitudes for $%
gg\rightarrow gg$ and $gg\rightarrow ggg$ processes (or called 22 and 23
processes in this paper), respectively. One useful observation is that the
right hand sides of Eqs. (\ref{eta1}) and (\ref{eta2}) correspond to
integrations over both sides of the Boltzmann equation, or equivalently, a
projection of the Boltzmann equation. It is certainly easier to solve the
projected equation than the Boltzmann equation itself. However, there would
be an infinite number of solutions satisfying the projected equation, even
though the true solution is unique, corresponding to that which gives the
largest $\eta $ (see e.g. \cite{Arnold:2003zc}). This makes solving $\eta $
a variational problem.

To solve for $B(p)$, we assume it to be a smooth function which can be
expanded in a set of orthogonal polynomials,
\begin{equation}
B(z)=z^{y}\sum_{r=0}^{\infty }b_{r}B^{(r)}(z),  \label{Bz}
\end{equation}
where $z=\beta |p|$, $B^{(r)}(z)$ is a polynomial up to $z^{r}$ and the
overall factor $z^{y}$ will be chosen by trial and error to get the fastest
convergence \cite{Dobado:2001jf}. The $B^{(r)}(z)$ polynomials can be
constructed using the condition
\begin{equation}
\int \frac{d^{3}\mathbf{p}}{(2\pi )^{3}}f_{p}^{0}(1+f_{p}^{0})|\mathbf{p}%
|z^{y}B^{(r)}(z)B^{(s)}(z)=T^{4}\delta _{rs}\text{ .}  \label{Ortho}
\end{equation}%
One can solve the coefficients $b_{r}$ by equating Eqs. (\ref{eta1}) and (%
\ref{eta2}). Then, $\eta $ is just proportional to $b_{0}$ according to Eqs.
(\ref{eta1}) and (\ref{Ortho}). For practical reasons, one uses the
approximation $B(z)=z^{y}\sum_{r=0}^{n-1}b_{r}B^{(r)}(z)$ where $n$ is a
finite, positive integer. It can be proved that $\eta $ is an increasing
function of $n$. Thus, one can systematically approach the true value of $%
\eta $. For $y=1$, the series converges rapidly. From $n=2$ to $3$, $\eta $
only changes by $\sim 1\%$.

In vacuum, $|M_{12\rightarrow 34}|^{2}=(12\pi \alpha _{s})^{2}\left(
3-tu/s^{2}-su/t^{2}-st/u^{2}\right) /2$ (see e.g. \cite{Ellis:1985er}). In
medium, $s=O(T^{2})$. The most singular part of $|M_{12\rightarrow 34}|^{2}$
comes from the colinear region (i.e. either $t\approx 0$\ or $u\approx 0$)
which can be regularized by the Hard-Thermal-Loop (HTL) dressed propagators
for gluons. However XG only used the Debye mass $m_{D}=(4\pi \alpha
_{s})^{1/2}T$ as the regulator just as done in Ref. \cite{Biro:1993qt}, so
for the sake of comparison between AMY and XG, we also use $m_D$ as the
regulator for soft and collinear divergences in this paper. We will use the
HTL gluon propagators in one of our future study. Thus, we consider the near
collinear approximation
\begin{equation}
|M_{12\rightarrow 34}|^{2}\approx -\frac{(12\pi \alpha _{s})^{2}}{2}\left.
\left( su/t^{2}+st/u^{2}\right) \right\vert _{\substack{ t\approx 0  \\
\text{or\ }u\approx 0}}\ .
\end{equation}%
In the center-of-mass (CM) frame, we can use the crossed symmetry between
the $u$- and $t$-channels and just use two times of the forward angle, $t$%
-channel contribution for the sum of the forward ($t$-channel) and backward (%
$u$-channel) angle contributions%
\begin{equation}
|M_{12\rightarrow 34}|^{2}\approx -(12\pi \alpha _{s})^{2}\left.
su/t^{2}\right\vert _{t\approx 0}\approx (12\pi \alpha _{s})^{2}\left. \frac{%
s^{2}}{(\mathbf{q}_{T}^{2}+m_{D}^{2})^{2}}\right\vert _{q^{2}\approx 0}\ ,
\label{M22}
\end{equation}%
where $q_{T}$\ is the transverse (with respect to $p_{1}$) component of $%
q=p_{2}-p_{4}$. Because small $q_{T}$\ could also come from large $%
\left\vert q^{2}\right\vert $\ through the $u$-channel, it is important to
note that when using Eq. (\ref{M22}) to calculate the collisional integral,
we only pick up the near forward scattering (around $t=q^{2}\approx 0$) to
avoid double counting.

%One might think that the
%above equation only includes the $t\approx 0$ region but not the
%$u\approx 0$ one. This is not the case since both the $t\approx 0$
%and $u\approx 0$ regions give $q_{T}\approx 0$ in the CM frame.

For the $gg\rightarrow ggg$ process, we will take the approximation that the
bremsstrahlung gluon is very soft (zero rapidity limit) and $\sqrt{s}$ is
much bigger than all transverse momenta. Then the exact result of Ref. \cite%
{Ellis:1985er} reduces to the Gunion-Bertsch formula \cite{Gunion:1981qs},
\begin{equation}
|M_{12\rightarrow 345}|^{2}\approx \sum_{\mathrm{perm}(3,4,5)}\frac{(12\pi
\alpha _{s})^{2}}{2}\left. \frac{s^{2}}{(\mathbf{q}_{T}^{2}+m_{D}^{2})^{2}}%
\frac{48\pi \alpha _{s}\mathbf{q}_{T}^{2}}{\mathbf{k}_{T}^{2}[(\mathbf{q}%
_{T}-\mathbf{k}_{T})^{2}+m_{D}^{2}]}\right\vert _{q^{2}\approx 0}\ ,
\label{eq:matrix-e}
\end{equation}%
where we have inserted the regulator $m_{D}^{2}$ as in Ref. \cite%
{Biro:1993qt}. Here $\mathbf{k}_{T}$\ is the transverse component of the
bremsstrahlung gluon momentum ($p_{5}$) and $\mathbf{q}_{T}$\ is still the
transverse component of $q=p_{2}-p_{4}$. The three final state gluons are
identical particles. Thus, there are 3! permutations of $(p_{3},p_{4},p_{5})$%
, each gives the same contribution. As explained above in the 22 case, we
need to be careful about using the $q_{T}$\ variable. Small $q_{T}$\ could
mean either the forward ( $t\approx 0$\ ) or backward ($u\approx 0$)
scattering. In the convention adopted for Eq. (\ref{eq:matrix-e}), one can
only pick up the near forward scattering (around $t=q^{2}\approx 0$) but not
the backward scattering otherwise double counting will happen. Our $%
|M_{12\rightarrow 345}|^{2}$\ is derived from the exact result of Ref. \cite%
{Ellis:1985er}, where Lorentz invariant Mandelstam variables are used so
there is no this ambiguity, after taking the soft bremsstrahlung limit. Eq. (%
\ref{eq:matrix-e}) is also consistent with the Gunion-Bertsch formula \cite%
{Gunion:1981qs} as explicitly demonstrated in App. \ref{sec-sig23gb}.
Effectively, the above treatment of collisional integrals leads to a factor
2 difference in the $gg\rightarrow ggg$ contribution to $\eta $ from that of
XG \cite{Xu} (ours is one half of XG's).

Naively the $gg\rightarrow gg$ collision rate is $\propto \int
dq_{T}^{2}|M_{12\rightarrow 34}|^{2}=O(\alpha _{s})$ and the $gg\rightarrow
ggg$ rate is $\propto \int dq_{T}^{2}dk_{T}^{2}|M_{12\rightarrow
345}|^{2}=O(\alpha _{s}^{2})$ (as will be discussed below, $k_{T}^{2}$ has
an $O(\alpha _{s})$ infrared (IR) cut-off). Thus, the $gg\rightarrow gg$
process seems more important than $gg\leftrightarrow ggg$. However, this is
incorrect. In $gg\rightarrow gg$, the amplitude is the largest in the
forward and backward scatterings. But there is no contribution to $\eta $ in
these cases since there is no momentum redistribution. Mathematically, we
have the additional suppression factor $%
[B_{ij}(p_{4})+B_{ij}(p_{3})-B_{ij}(p_{2})-B_{ij}(p_{1})]^{2}\simeq
O(q_{T}^{2})$ in Eq. (\ref{eta2}), while no similar suppression in $%
gg\leftrightarrow ggg$. Thus, the $gg\rightarrow gg$ collision rate is
proportional to $\int dq_{T}^{2}|M_{12\rightarrow 34}|^{2}q_{T}^{2}=O(\alpha
_{s}^{2}\log \alpha _{s})$, which is of the same order as $O(\alpha
_{s}^{2}) $ of $gg\leftrightarrow ggg$, up to a logarithm \cite%
{Arnold:2000dr,Arnold:2003zc}.

This power counting can be used to argue that other processes such as $%
ggg\rightarrow ggg$\ and $gg\rightarrow gggg$\ (called 33 and 24 processes)
are higher order under the assumption that the most important contribution
to $\eta $\ comes from the configurations with at most two hard gluons in
the initial or the final states. Under this momentum configuration, one
observes in Eq. (\ref{eq:matrix-e})\ that adding a soft gluon to the 22
process yields a factorizable form for the 23 matrix element squared.
Schematically,
\begin{equation}
\frac{|M_{23}|^{2}}{|M_{22}|^{2}}\simeq O(\alpha _{s}p_{T}^{-2}),
\end{equation}%
where $p_{T}$\ denotes the small momentum scale with $p_{T}\simeq
O(q_{T})\simeq O(k_{T})$. Analogously, adding a soft gluon to the 23 process
yields
\begin{equation}
\frac{|M_{33(24)}|^{2}}{|M_{23}|^{2}}\simeq O(\alpha _{s}p_{T}^{-2}).
\end{equation}%
Thus, the 33(24) collision rate is smaller than that of 23 by a factor of $%
\int dp_{T}^{2}|M_{33(24)}|^{2}/|M_{23}|^{2}=O(\alpha _{s}\log \alpha _{s})$%
. This argument can be generalized to other processes as well. Thus, 22 and
23 are the only processes in the LO under this assumption.

The phase space of the 3-gluon state plays an important role in the
collisional integral in Eq. (\ref{eta2}) for $gg\leftrightarrow ggg$, which
is controlled by the delta-functions for energy-momentum conservation. Since
we use the Gunion-Bertsch formula, Eq. (\ref{eq:matrix-e}), which is valid
for soft gluon bremsstrahlung, it is consistent to apply the same condition
for energy-momentum configuration of the 3-gluon state. This can be done by
neglecting the recoil effect due to the soft gluon bremsstrahlung, i.e.
neglecting the momentum of the soft gluon inside the delta-functions as is
done in App. \ref{sec-sig23gb}. Therefore, the phase space for the two near
collinear gluons in 3-gluon state is 2-body-like. Additionally the exact
phase space is 3-body-like if the momentum of the soft gluon is kept and
treated in equal footing as the other gluons in the delta-functions. We will
see that using the 3-body-like or 2-body-like phase space makes a
significant difference in the shear viscosity.

\section{Leading-Log Result}

In the leading-log (LL) approximation, one just needs to focus on the small $%
q_{T}$ contribution from the $gg\rightarrow gg$ process. After performing
the small $q_{T}$ expansion to Eq. (\ref{eta2}), we obtain ($g^{2}=4\pi
\alpha _{s}$)

\begin{equation}
\eta _{LL}\simeq 27.1\frac{T^{3}}{g^{4}\ln (1/g)},  \label{LL}
\end{equation}%
which agrees with that of \cite{Arnold:2000dr} very well. Using the entropy
density for non-interacting gluons, $s=N_{g}\frac{2\pi ^{2}}{45}T^{3}$, we
obtain
\begin{equation}
\frac{\eta _{LL}}{s}\simeq \frac{3.9}{g^{4}\ln (1/g)}.  \label{LLs}
\end{equation}%
This will be used to check our numerical result later. In contrast, we take
the Boltzmann gas approximation ($f_{p}^{0}=e^{-E_{p}/T}$) used by XG, we
get $\eta _{LL}\simeq 44.7T^{3}g^{-4}\ln ^{-1}(1/g)$ and $s=N_{g}\frac{4}{%
\pi ^{2}}T^{3}$, which would yield $\eta _{LL}/s\simeq 6.9g^{-4}\ln
^{-1}(1/g)$. Thus, the error from taking the Boltzmann gas approximation for
the LL result of $\eta /s$ is $\sim $80\%, where $\sim $70\% comes from $%
\eta $\ and $\sim $10\% comes from $s$. This suggests that the quantum
nature of gluons could play an important role on transport coefficients,
even though they might not be important for thermodynamic quantities. In
weak coupling regime, e.g. $\alpha _{s}=10^{-3}$, the XG result in \cite%
{Xu:2007ns} gives $\eta _{22}/s\approx 5.6\times 10^{3}$ while the LL result
gives $\eta _{22}/s\approx 2\times 10^{4}$, which shows a factor 4
difference. But the difference from the LL result can be narrowed in
Israel-Stewart theory \cite{El:2008yy}.

\section{Treatment of $gg\leftrightarrow ggg$}

As mentioned above, both $gg\rightarrow gg$ and $gg\leftrightarrow ggg$ are
needed to compute $\eta $ to the leading order ($O(\alpha _{s}^{-2})$). For
the treatment of the 23 process, we consider three cases, (a) with the
3-body-like phase space for three gluons and with the LPM effect as the
cutoff for the soft gluon; (b) with the 3-body-like phase space and but with
$m_D$ as the regulator for the soft gluon; (c) with the 2-body-like phase
space and with $m_D$ as the regulator for the soft gluon.

In case (a), the scale of the $k_{T}$ cut-off is set by the
Landau-Pomeranchuk-Migdal (LPM) effect, as in Refs. \cite{Xu:2007ns} and
\cite{Arnold:2003zc}. Ref. \cite{Gyulassy:1991xb} gives an intuitive
explanation of the LPM effect: for the bremsstrahlung gluon with transverse
momentum $|k_{T}|$, the mother gluon has a transverse momentum uncertainty $%
\sim |k_{T}|$ and a size uncertainty $\sim 1/|k_{T}|$. It takes the
bremsstrahlung gluon the formation time $t\sim 1/|k_{T}|v_{T}\sim
E_{k}/|k_{T}|^{2}$ to fly far enough from the mother gluon to be resolved as
a radiation. But if the formation time is longer than the mean free path $%
l_{mfp}\approx O(\alpha _{s}^{-1})$, then the radiation is incomplete and it
would be resolved as $gg\rightarrow gg$ instead of $gg\rightarrow ggg$.
Thus, the resolution scale is set by $t\leq l_{mfp}$. This yields the
condition $|k_{T}|^{2}\geq E_{k}/l_{mfp}\approx O(\alpha _{s})$\ which is
confirmed through carefully derivations in Ref. \cite{Wang:1994fx}.

Here the mean free path $l_{mfp}$ is given by the collision rate $R\simeq
1/l_{mfp}$ which sets the scale of the LPM effect is computed via the
detailed balance rate. After integration, the Boltzmann equation of Eq.(\ref%
{Boltzmann}) can be written as
\begin{equation}
\frac{dn}{dt}=n\left( R_{gain}-R_{loss}\right) ,
\end{equation}%
where we have used $n=\int \frac{d^{3}\mathbf{p}}{(2\pi )^{3}}f_{p}$. Then
the collision rate is the detailed balance rate in thermal equilibrium,
\begin{equation}
R\equiv R_{gain}^{equil.}=R_{22}+R_{23}+R_{32}\text{,}
\end{equation}%
where
\begin{eqnarray}
R_{22} &=&\frac{N_{g}}{2n}\int \prod_{i=1}^{4}\frac{d^{3}p_{i}}{\left( 2\pi
\right) ^{3}2E_{i}}\left\vert M_{12;34}\right\vert ^{2}(2\pi )^{4}\delta
^{4}(p_{1}+p_{2}-p_{3}-p_{4})  \notag \\
&&\times f_{1}^{0}f_{2}^{0}\left( 1+f_{3}^{0}\right) \left(
1+f_{4}^{0}\right) ,  \notag \\
R_{23} &=&\frac{N_{g}}{6n}\int \prod_{i=1}^{5}\frac{d^{3}p_{i}}{\left( 2\pi
\right) ^{3}2E_{i}}\left\vert M_{12;345}\right\vert ^{2}(2\pi )^{4}\delta
^{4}(p_{1}+p_{2}-p_{3}-p_{4}-p_{5})  \notag \\
&&\times f_{1}^{0}f_{2}^{0}\left( 1+f_{3}^{0}\right) \left(
1+f_{4}^{0}\right) \left( 1+f_{5}^{0}\right) ,  \notag \\
R_{32} &=&\frac{3}{2}R_{23}.  \label{rate}
\end{eqnarray}%
Note that our definition of $R$\ is the same as that of XG. The phase space
for three gluons is 3-body-like in $R_{23}$. And\textbf{, }as mentioned
above, only near forward scattering (around $t=q^{2}\approx 0$) is included
to avoid double counting (see App. \ref{sec-sig23gb}), which gives an
additional factor $1/2$ compared to XG \cite{Xu:2007ns}. We have computed $R$
self-consistently since $R_{23}$ also depends on $R$. Our $R_{22}$ and $%
R_{23}$, together with the results for the Boltzmann gas approximation ($%
f_{p}^{0}=e^{-E_{p}/T}$, $1+f_{p}^{0}\rightarrow 1$), are shown in Fig. 1.
Our $R_{22}$, which uses Bose-Einstein (BE) statistics, is close to the
Boltzmann gas result. Our $R_{23}$, however, gets an enhancement for $\alpha
_{s}\lesssim 0.04$ from the enhancement factor $\left( 1+f_{5}^{0}\right) $
which is inversely proportional to the soft gluon's bremsstrahlung energy.
This enhancement in $R$ makes the $gg\leftrightarrow ggg$ contribution to $%
\eta $ smaller in the BE case than the Boltzmann gas. The enhancement
disappears at higher $\alpha _{s}$ where $R/T$, and hence the IR cut-off,
becomes bigger.

%------------------------------------------------------
\begin{figure}[tbp]
\begin{center}
\includegraphics[scale=0.6]{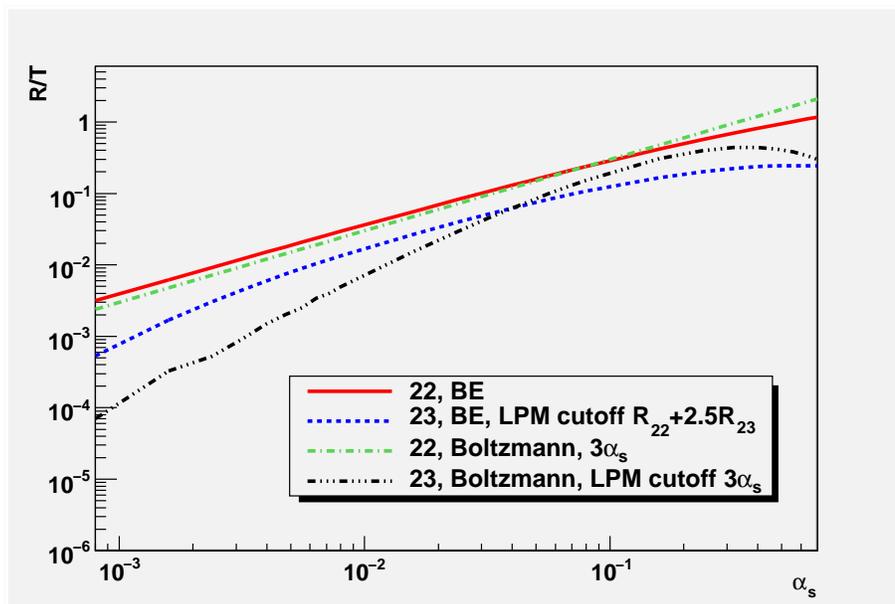}
\end{center}
\caption{(color online) $R_{22}$ and $R_{23}$ of Eq.(\protect\ref{rate})
shown as functions of $\protect\alpha _{s}$ for BE and Boltzmann gas. }
\label{ratefig}
\end{figure}
%------------------------------------------------------

In case (b) and (c) , we introduce an IR cut-off $m_{D}$ by replacing the $%
1/k_{T}^{2}$ factor in Eq. (\ref{eq:matrix-e}) with $1/(k_{T}^{2}+m_{D}^{2})$%
. Thus, $m_{D}$ not only screens the intermediate states but also the
external states. This is motivated by demanding the optical theorem to be
valid in the medium, even though it need not be the case when the system
exchanges particles from a thermal bath. Thus, if the propagator in the loop
is screened, then the bremsstrahlung gluon is also screened. This very naive
treatment gives $|k_{T}|^{2}\gtrsim m_{D}^{2}=O(\alpha _{s})$, which is
consistent with the first treatment in the $\alpha _{s}$ counting.

\section{Numerical Results and Discussions}

We show in Fig. 2 the comparison between $\eta $ computed with $%
gg\rightarrow gg$ alone (denoted as $\eta _{22}$) and $\eta $ computed with $%
gg\rightarrow gg$ and $gg\leftrightarrow ggg$ (denoted as $\eta _{22+23}$)
\cite{statistic}. In computing the 23 contribution in case (a) and (b) with
the 3-body-like phase space for three gluons, we use different treatments of
$k_{T}$ cut-offs: in case (a) we use $R=R_{22}+2.5R_{23}$ as the cut-off,
where $R_{23}$ is self-consistently determined (the blue dashed line in Fig. %
\ref{ratefig}), while in case (b) we use $m_{D}$ as the regulator. For these
two cases we find that adding $gg\leftrightarrow ggg$ reduces $\eta $ by $%
\sim 30\%$ at $\alpha _{s}=10^{-3} $ where the contribution from $%
gg\leftrightarrow ggg$ is about 1/2 of that from $gg\leftrightarrow gg$. The
correction is the largest, $\sim 75\%$, at $\alpha _{s}=0.1$. This means the
$gg\leftrightarrow ggg$ contribution is about 3 times that of $gg\rightarrow
gg$. The behavior shown here is different from that of XG which shows $\eta
_{22+23}/\eta _{22}\sim 1/8\sim 12.5\%$, meaning that the $gg\leftrightarrow
ggg$ contribution is about 7 times as large as $gg\rightarrow gg$, for a
wide range of $\alpha _{s}$ ($\alpha _{s}=10^{-3}-0.7$). The difference
between our result and XG's is largely due to the factor 2 difference in
collisional integrals for the $gg\leftrightarrow ggg$ process and the BE
statistics versus the Boltzmann gas approximation used. But we do see the
dominance of $gg\leftrightarrow ggg $ over $gg\rightarrow gg$ when $\alpha
_{s}\gtrsim 2\times 10^{-3}$, as asserted by XG.

For case (c) with the 2-body-like phase space for three gluons the effect of
the 23 process is about 10-30\%, which is close to AMY's result in the whole
range of $\alpha _{s}$.\textbf{\ }Since our result changes dramatically
after imposing the soft bremsstrahlung approximation, it means this
approximation is not as good as previously expected. Thus, it is important
to go beyond this approximation to obtain an accurate $\eta $.

In Fig. 3, $\eta /s$ as a function of $\alpha _{s}$ is shown for different
cases: the LL result $\eta _{LL}/s$ of Eq. (\ref{LLs}), $\eta _{22}/s$, and $%
\eta _{22+23}/s$ for two different $k_{T}$ cut-offs for the
3-body-like phase space and that for the 2-body-like phase space.
When $\alpha _{s}\rightarrow 0$, all these curves should converge to
the LL result. But at $\alpha _{s}=10^{-3}$, we have $\ln
(1/g)=2.2$, which is not large enough to dominate the contribution.
This is the reason for the deviations of the numerical results from
the LL one in the current range of $\alpha _{s}$. However, the
agreement between the $\eta _{LL}/s$ and $\eta _{22}/s$ is a good
check to our numerical calculations which are carried out by the
Monte Carlo method for multi-dimensional integrations. The power of
$\alpha _{s}$ dependence of these curves are close to $(-2)$ as
expected. At $\alpha _{s}=0.1$, with both $k_{T}$ cut-offs for the
3-body-like phase space, the full result $\eta _{22+23}/s\simeq 1.0$
is between 2.7 of AMY \cite{footnote} and $0.5$ of XG. At $\alpha
_{s}=0.3$ and 0.6, we have $\eta _{22+23}/s\simeq 0.22$ and 0.15,
respectively, which are larger than 0.13 and 0.076 obtained by XG.
It is also interesting to note the good agreement using two
different cut-offs for the bremsstrahlung gluon momentum. For the
2-body-like phase space the correction from the 23 process is small
and $\eta _{22+23}/s\approx (70\%\sim 90\%)\eta _{22}/s$, which is
close to AMY's result.

%------------------------------------------------------
\begin{figure}[tbp]
\begin{center}
\includegraphics[scale=0.6]{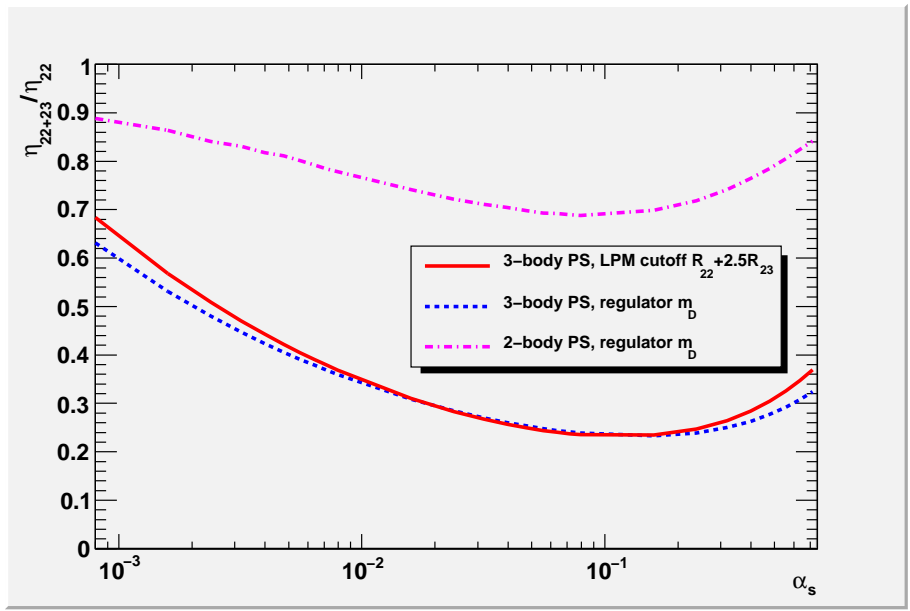}
\end{center}
\caption{(color online) $\protect\eta _{22+23}/\protect\eta _{22}$ shown as
a function of $\protect\alpha _{s}$ for the 3-body-like and 2-body-like
phase space (PS) of three gluons. There are two different treatments of the
cut-off of the bremsstrahlung gluon momentum $k_{T}$ for the 3-body-like
phase space. }
\end{figure}
%------------------------------------------------------

%------------------------------------------------------
\begin{figure}[tbp]
\begin{center}
\includegraphics[scale=0.6]{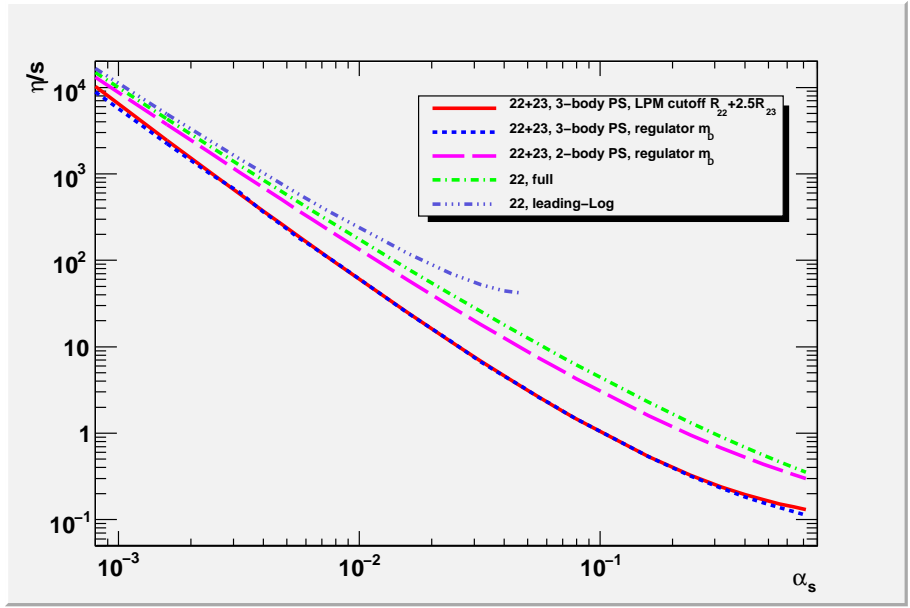}
\end{center}
\caption{(color online) $\protect\eta /s$ versus $\protect\alpha _{s}$ for
(a) the leading-log result in Eq. (\protect\ref{LL}), (b) the result of the
22 process only, the full result with 22+23 processes for the 3-body-like
phase space (PS) of three gluons where the $k_{T}$\ cut-off is set by (c) $%
m_{D}$ or (d) the LPM effect, and (e) the full result with 22+23 processes
for the 2-body-like phase space (PS).}
\end{figure}
%------------------------------------------------------

In summary, we have calculated the shear viscosity over entropy density $%
\eta /s$ of a gluon plasma in kinetic theory. Due to the suppressed
contribution to $\eta $\ in the $gg\rightarrow gg$\ forward scattering, the
gluon bremsstrahlung $gg\leftrightarrow ggg$ process also contributes at the
same order ($O(\alpha _{s}^{-2})$) in perturbative QCD. We find that the $%
gg\leftrightarrow ggg$ contribution becomes more important to $\eta $ than $%
gg\rightarrow gg$ for $\alpha _{s}\gtrsim 2\times 10^{-3}$ for the
3-body-like phase space for the three-gluons state. At $\alpha _{s}=0.1$, $%
\eta /s\simeq 1.0$ which is between 2.7 obtained by Arnold, Moore and Yaffe
\cite{Arnold:2003zc} and 0.5 obtained by Xu and Greiner \cite{Xu:2007ns}.
Our $\eta /s$ is about 2 times as large as that of Xu and Greiner for $%
\alpha _{s}\gtrsim 0.1$, largely due to the factor 2 difference in
collisional integrals for the $gg\leftrightarrow ggg$ process and the
Bose-Einstein statistics versus the Boltzmann gas approximation used. We
have observed that using $m_{D}$ as the regulator for transverse momentum of
the soft bremsstrahlung gluon agrees well with that using the rate as the
cut-off for the LPM effect in $\eta $ for the current range of $\alpha _{s}$%
. In dealing with the 23 process it is consistent to implement the soft
gluon condition in the energy-momentum configuration of the three-gluons
state that there is one soft gluon, which results in the 2-body-like phase
space for the three-gluons state, since we use the Gunion-Bertsch formula
for the 23 matrix element which is valid only for soft gluon bremsstrahlung.
In this case we obtain results close to AMY's. To test which is the correct
description for the phase space of three gluons in the 23 process, or in
other words, to test if the Gunion-Bertsch formula is still valid for
general 3-body-like momentum configurations, a further and comprehensive
study with the exact matrix element is needed.

\bigskip

Acknowledgement: JWC and QW thank Zhe Xu for clarifying the definition of
collisional rates used in the LPM effect and for many helpful discussions.
The authors also thank Carsten Greiner and Guy Moore for helpful comments.
JWC and KO are supported by the NSC and NCTS of R.O.C.. QW is supported in
part by the '100 talents' project of Chinese Academy of Sciences (CAS) and
by the National Natural Science Foundation of China (NSFC) under the grants
10675109 and 10735040. HD is supported by NSFC under the grant 10847149.

\appendix

\section{The cross section for 23 from Gunion-Bertsch formula}

\label{sec-sig23gb} In the center-of-mass frame of 1 and 2, the cross
section is written by,
\begin{eqnarray}
\sigma_{23} & = & \frac{1}{2s}\frac{1}{3!}\int\prod_{i=3}^{5}\frac{d^{3}k_{i}%
}{(2\pi)^{3}2E_{i}} |M_{12;345}|^{2}(2\pi)^{4}%
\delta^{4}(k_{1}+k_{2}-k_{3}-k_{4}-k_{5})  \notag \\
& = & \frac{27}{\pi^{2}}\alpha_{s}^{3}\int d^{3}k_{3} \frac{1}{(\mathbf{q}%
_{T}^{2}+m_{D}^{2})^{2}}\delta(E_{1}+E_{2}-E_{3}-E_{4}) \int d^{2}k_{T}dy%
\frac{\mathbf{q}_{T}^{2}}{\mathbf{k}_{T}^{2}[(\mathbf{q}_{T}-\mathbf{k}%
_{T})^{2}+m_{D}^{2}]}  \notag \\
& = & \frac{27}{\pi^{2}}\alpha_{s}^{3} \int d^{2}q_{T}\frac{1}{(\mathbf{q}%
_{T}^{2}+m_{D}^{2})^{2}} \int d^{2}k_{T}dy \frac{\mathbf{q}_{T}^{2}}{\mathbf{%
k}_{T}^{2}[(\mathbf{q}_{T}-\mathbf{k}_{T})^{2}+m_{D}^{2}]}.
\label{eq:sig23-GB}
\end{eqnarray}
Since we use the Gunion-Bertsch formula for soft gluon bremsstrahlung, we
assume the 5th gluon is soft, so we made the approximation in the second
equality of Eq. (\ref{eq:sig23-GB}),
\begin{equation}
\delta^{4}(k_{1}+k_{2}-k_{3}-k_{4}-k_{5})\approx%
\delta^{4}(k_{1}+k_{2}-k_{3}-k_{4}),
\end{equation}
which means the phase space is dominated by the 22 process. We also used $%
E_{3}=E_{4}=E_{1}=E_{2}=\sqrt{s}/2$ and
\begin{eqnarray}
\int d^{3}k_{3}\delta(E_{1}+E_{2}-E_{3}-E_{4}) & = & \frac{1}{2}\int
d^{3}q\delta(E_{1}-E_{3})  \notag \\
& = & \frac{1}{2}\int d^{2}q_{T}dq_{z}\delta(E_{1}-\sqrt{%
(E_{1}+q_{z})^{2}+q_{T}^{2}})  \notag \\
& = & \int d^{2}q_{T}\frac{E_{1}}{\sqrt{E_{1}^{2}-q_{T}^{2}}}\approx\int
d^{2}q_{T}
\end{eqnarray}
where $\mathbf{k}_{3}=\mathbf{k}_{1}+\mathbf{q}$. Note that a factor of 2 is
given from the two roots for $q_{z}$ in the equation $E_{1}=\sqrt{%
(E_{1}+q_{z})^{2}+q_{T}^{2}}$, i.e. $q_{z}=-E_{1}\pm\sqrt{E_{1}^{2}-q_{T}^{2}%
}$ which correspond to forward and backward solution $q_{z}=-\sqrt{s},0$ or $%
t=-s,0$ at $q_{T}=0$. Eq. (\ref{eq:sig23-GB}) is 2 times as large as that
derived in Ref. \cite{Biro:1993qt}. One has to choose the forward scattering
and get the factor 1/2,
\begin{equation}
\int_{forward}d^{3}k_{3}\delta(E_{1}+E_{2}-E_{3}-E_{4})=\frac{1}{2}\int
d^{2}q_{T}.
\end{equation}
Then the differential cross section from Eq. (\ref{eq:sig23-GB}) becomes
\begin{equation}
\frac{d\sigma_{23}}{d^{2}q_{T}d^{2}k_{T}dy}=\frac{27}{2\pi^{2}}%
\alpha_{s}^{3} \frac{1}{(\mathbf{q}_{T}^{2}+m_{D}^{2})^{2}} \frac{\mathbf{q}%
_{T}^{2}}{\mathbf{k}_{T}^{2}[(\mathbf{q}_{T}-\mathbf{k}_{T})^{2}+m_{D}^{2}]},
\end{equation}
which reproduces the result in \cite{Biro:1993qt}.

%%%%%%%%%%%%%%%%%%%%%%%%%%%%%%%%%%%%%%%%%%%%%%%%%%%%%%%%%%%%%%


\begin{thebibliography}{99}
\bibitem{statistic} We use Monte-Carlo integration method to carry out
multi-dimensional phase space integrals. We calculate 27 points of $\alpha_s$%
, $\alpha_s=0.01n/(4\pi),0.1n/(4\pi),n/(4\pi)$ with $n=1,2,...,9$. The
number of event samples for each data point is $10^{10}$. The fluctuations
for $\eta _{22+23}$ with LPM cutoff for very small coupling constant, $%
\alpha_s \lesssim 3\times 10^{-3}$, are quite large. The $\alpha_s \lesssim
3\times 10^{-3}$ part of the red solid curves in Fig. 1 and Fig. 2 are
fitted results and subject to change for larger numbers of event samples.
The results outside this range converge very well for current event samples.
\end{thebibliography}
\end{document}